\documentclass[prb,twocolumn,preprintnumbers,amsmath,amssymb,showpacs,superscriptaddress,floatfix]{revtex4}

\usepackage{graphicx}
\usepackage{dcolumn}
\usepackage{bm}
\usepackage{nicefrac}

\begin{document}


\title{Static and fluctuating stripe order observed by resonant soft x-ray diffraction in La$_{1.8}$Sr$_{0.2}$NiO$_{4}$ }

\author{J. Schlappa}
\affiliation{II. Physikalisches Institut, Universit{\"a}t zu
K{\"o}ln, Z{\"u}lpicher Str.~77, D-50937 K{\"o}ln, Germany}
\author{C.~F. Chang}
\affiliation{II. Physikalisches Institut, Universit{\"a}t zu K{\"o}ln,
Z{\"u}lpicher Str.~77, D-50937 K{\"o}ln, Germany}
\author{E. Schierle}
\affiliation{Institut f{\"ur} Experimentalphysik, Freie Universit{\"a}t Berlin,
Arnimallee 14, D-14195 Berlin, Germany}
\affiliation{Helmholtz Centre Berlin for Materials and Energy,
Albert-Einstein-Str. 15, 12489 Berlin, Germany}
\author{A.~Tanaka}
 \affiliation{Department of Quantum Matter, ADSM, Hiroshima University,
 Higashi-Hiroshima 739-8530, Japan}
\author{R. Feyerherm}
\affiliation{Helmholtz Centre Berlin for Materials and Energy,
Albert-Einstein-Str. 15, 12489 Berlin, Germany}
\author{Z. Hu}
\affiliation{II. Physikalisches Institut, Universit{\"a}t zu K{\"o}ln,
Z{\"u}lpicher Str.~77, D-50937 K{\"o}ln, Germany}
\author{H. Ott}
\affiliation{II. Physikalisches Institut, Universit{\"a}t zu K{\"o}ln,
Z{\"u}lpicher Str.~77, D-50937 K{\"o}ln, Germany}
\author{O. Friedt}
\affiliation{II. Physikalisches Institut, Universit{\"a}t zu K{\"o}ln,
Z{\"u}lpicher Str.~77, D-50937 K{\"o}ln, Germany}
\affiliation{Laboratoire L\'eon Brillouin, CEA-Saclay, 91191 Gif-sur-Yvette Cedex, France}
\author{E. Dudzik}
\affiliation{Helmholtz Centre Berlin for Materials and Energy,
Albert-Einstein-Str. 15, 12489 Berlin, Germany}
\author{H.-H. Hung}
\affiliation{National Hsinchu University of Education, Hsinchu, Taiwan}
\author{M. Benomar}
\affiliation{II. Physikalisches Institut, Universit{\"a}t zu K{\"o}ln,
Z{\"u}lpicher Str.~77, D-50937 K{\"o}ln, Germany}
\author{M. Braden}
\affiliation{II. Physikalisches Institut, Universit{\"a}t zu K{\"o}ln,
Z{\"u}lpicher Str.~77, D-50937 K{\"o}ln, Germany}
\author{L. H. Tjeng}
\affiliation{II. Physikalisches Institut, Universit{\"a}t zu K{\"o}ln,
Z{\"u}lpicher Str.~77, D-50937 K{\"o}ln, Germany}%
\author{C. Sch{\"u}{\ss}ler-Langeheine}%
\email [Corresponding author: ]{schuessler@ph2.uni-koeln.de}
\affiliation{II. Physikalisches Institut, Universit{\"a}t zu K{\"o}ln,
Z{\"u}lpicher Str.~77, D-50937 K{\"o}ln, Germany}

\date{\today}

\begin{abstract}
We studied the stripe phase of La$_{1.8}$Sr$_{0.2}$NiO$_{4}$
using neutron diffraction, resonant soft x-ray diffraction (RSXD)
at the Ni $L_{2,3}$ edges, and resonant x-ray diffraction (RXD) at
the Ni $K$ threshold. Differences in the $q$-space resolution of
the different techniques have to be taken into account for a
proper evaluation of diffraction intensities associated with the
spin and charge order superstructures. We find that in the RSXD
experiment the spin and charge order peaks show the same
temperature dependence. In the neutron experiment by contrast,
the spin and charge signals follow quite different temperature
behaviors. We infer that fluctuating magnetic order contributes
considerably to the magnetic RSXD signal and we suggest that this
result may open an interesting experimental approach to search
for fluctuating order in other systems by comparing RSXD and
neutron diffraction data.
\end{abstract}


\maketitle

Sr-doped $\mathrm {La_2NiO_4}$ (LSNO) and $\mathrm {La_2CuO_4}$
(LSCO) form at low temperatures an ordered phase of charge and
spin degrees of freedom, the so called stripes. This phase
consists of one-dimensional hole rich lines, which form antiphase
domain walls for the antiferromagnetic background of the
hole-poor regions.\cite{zaanen:89a,emery:93a,tranquada:95b} The
stripe phase has attracted considerable interest because of its
possible relation to cuprate
superconductivity.\cite{orenstein:00a} While in LSNO static
stripe order is found at low temperatures, LSCO exhibits
fluctuating stripe order which can be rendered static by
additional doping.\cite{yamada:98a} From neutron diffraction it
is known that in LSNO static spin order (SO) sets in at a N\'eel
temperature $T_N$ well below the charge ordering (CO) temperature
$T_{CO}$.
\cite{tranquada:95a,tranquada:96a,lee:97a,yoshizawa:00a,lee:01a}
Yet, above $T_N$, fluctuating magnetic correlations can be found
at finite energies.\cite{lee:02a,sachan:95a}

Most experimental results concerning the structure of the stripe
phase are based on neutron diffraction experiments, which are
sensitive to magnetic order and to the lattice distortion caused
by the charge ordering. In a neutron scattering experiment one
can readily distinguish between static and fluctuating magnetic
ordering by energy analysis of the scattered neutrons. Static
order yields an elastic signal, while magnetic fluctuations lead
to a finite energy change of the scattered neutrons.

Recently resonant soft x-ray diffraction (RSXD) has been shown to be
another powerful experimental tool to study stripes or similar
order phenomena, because of its high sensitivity to spatial
modulations of the electronic state and its high magnetic
scattering contrast
\cite{abbamonte:02a,wilkins:03a,dhesi:04a,thomas:04a,schuessler:05a}.
It is therefore well suited to study both the charge and the
magnetic signal from a stripe system. We have used RSXD before to
identify the \textit{electronic} character of the stripe phase in
La$_{1.8}$Sr$_{0.2}$NiO$_{4}$.\cite{schuessler:05a} Now we would
like to address the question of the temperature dependence of SO
and CO in this system. Earlier conventional x-rays experiments
from CO in LSNO found an interesting discrepancy to the neutron
findings for the low temperature behavior: while the neutron CO
signal increases monotonically upon cooling and eventually
saturates,
\cite{hayden:92a,sachan:95a,tranquada:95a,tranquada:96a,lee:97a,lee:01a}
conventional x-rays found a maximum intensity at intermediate
temperatures and a decay of the CO signal upon cooling.
\cite{du:00a,hatton:02a,hatton:02b,ghazi:04a,spencer:05a} This
effect has been discussed in terms of a surface effect or of a
difference in the signals probed by neutron and x-ray
diffraction, \cite{spencer:05a} but no conclusive explanation has
been given so far.

Here we carried out a comprehensive study on
La$_{1.8}$Sr$_{0.2}$NiO$_{4}$ in which we compare the temperature
dependence of the SO and CO signals, and in which we use neutron
diffraction, RSXD at the Ni $L_{2,3}$ edges, and resonant x-ray
diffraction (RXD) at the $K$ threshold as techniques.

A La$_{1.8}$Sr$_{0.2}$NiO$_{4}$ single crystal was grown at the
University of Cologne using the traveling solvent method. The
crystalline quality was checked by x-ray diffraction and was found
to be very good with a rocking width of $0.01^\circ$ (FWHM) of the
(204) reflection observed using resonant x-ray diffraction at the
Ni-$K$-edge. Two pieces were cut and polished, one with a (101)
surface orientation and one with (103). We refer to the commonly
used orthorhombic unit cell with $a \approx b = 5.38$~{\AA} and
$c = 12.55$~{\AA} (space group $Fmmm$). In this setting the
direction of modulation of the electronic state in the NiO$_2$
planes is along $a$ and the stripes extend along $b$. Because of
twinning, domains with the stripe pattern rotated by 90 degrees
around $c$ exist as well.

\begin{figure}[t]
\includegraphics[scale=1,clip,width=8.0cm]{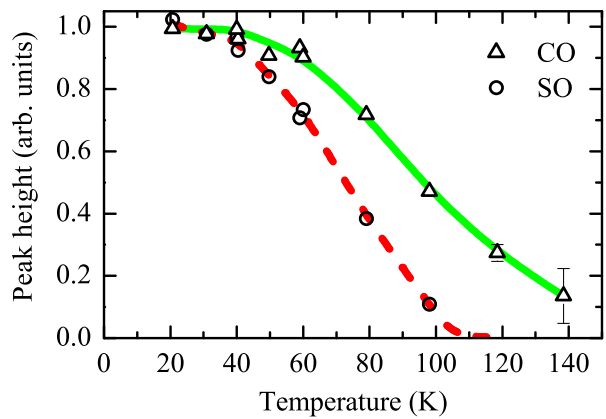}
\caption{(Color online) Temperature dependence of the peak height
of the $(4 - 2 \epsilon, 0, 1)$ charge-order peak (triangles) and
$(1-\epsilon, 0 , 1)$ spin-order peak (circles) as probed by
elastic neutron diffraction. The lines are guides to the eyes.}
\label{neutron}
\end{figure}

Neutron diffraction experiments were carried out at the Orph\'{e}e
reactor, diffractometer 3T.1 using a larger piece of the same
single crystal. We probed the elastic signal from the
$(1-\epsilon, 0 , 1)$ spin order and $(4 - 2 \epsilon, 0, 1)$
charge order peaks with $\epsilon$ being the incommensurability
parameter describing the spacing between stripes. The temperature dependence of the heights of both peaks is
presented in Fig.~\ref{neutron}. We found that the elastic
magnetic signal (circles) decays at about 105 K while the elastic
charge-order signal (triangles) is still visible at 140 K. The
behavior of the two signals is very similar to the one Sachan
\textit{et al.} found for the elastic signals from their sample
[Fig.~2(a) of Ref.~\onlinecite{sachan:95a}].

The RSXD experiments were performed from the (101) sample at the
soft x-ray beamlines U49/2-PGM1 and UE52-SGM of BESSY, using the
two-circle UHV diffractometer, designed at the Freie
Universit\"at Berlin. The scattering geometry and sample
orientation is described in Ref.~\onlinecite{schuessler:05a}. The
photon polarization was linear and could be switched from
parallel to the diffraction plane ($\pi$-polarization) to
perpendicular ($\sigma$). RXD experiments at the Ni-$K$ ($1s
\rightarrow 4p$) threshold at 8358 eV were performed from the
(103) sample at the \textit{MAGS}-beamline at BESSY.

\begin{figure}[t]
\includegraphics[width=8.0cm]{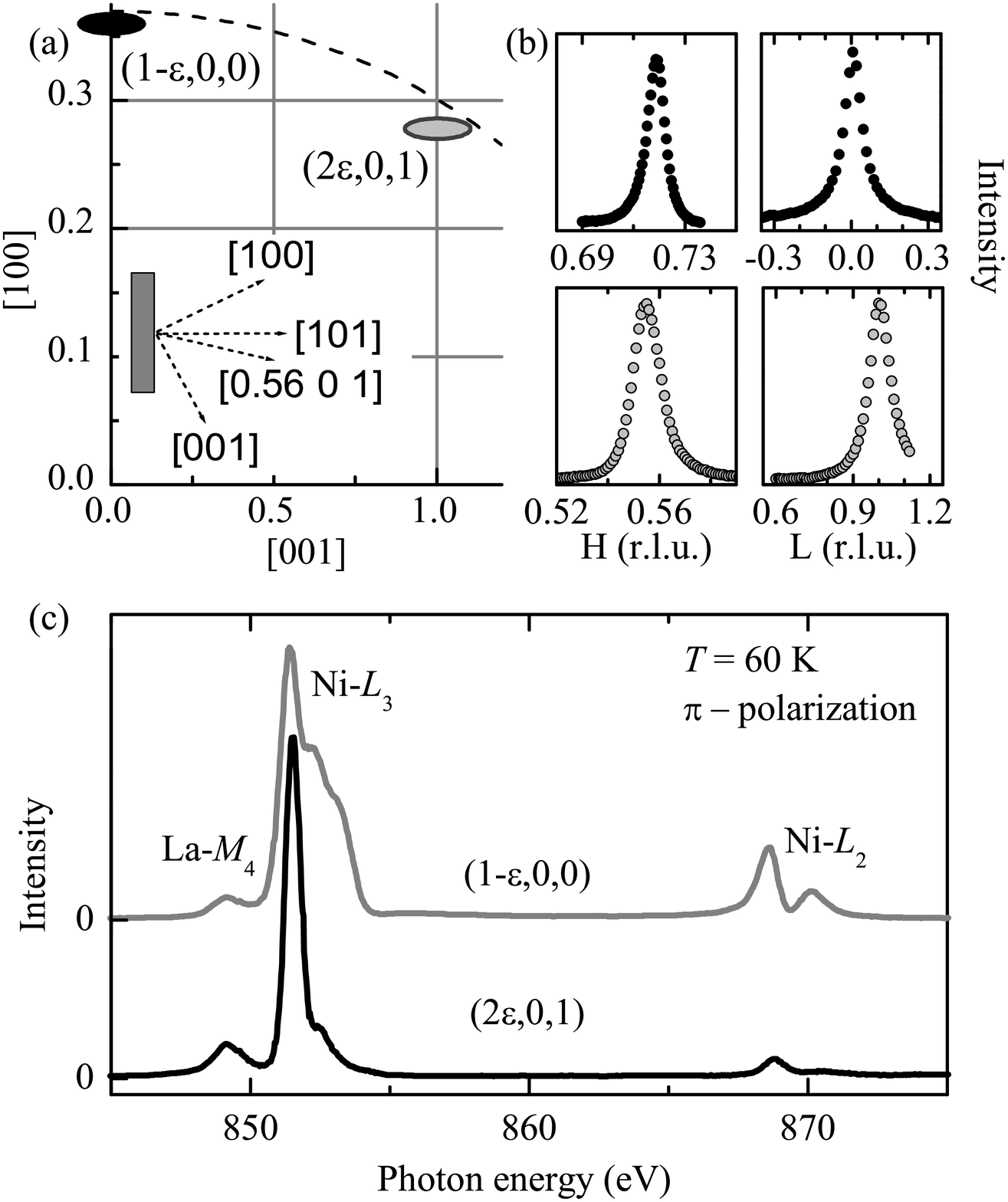}
\caption{(a) Position of the investigated superstructure peaks
in k-space and (b) scans along the H- and L-direction (right)
recorded at the Ni $L_{2,3}$ resonance with $\pi$-polarized light.
Black symbol: $(1-\epsilon,0,0)$ - spin order peak, gray symbol:
$(2\epsilon,0,1)$ - charge order peak. The dashed line in panel
(a) denotes the maximum possible momentum transfer at the Fe $L_3$
resonance. The inset shows the crystal directions in the scattering
plane of the soft x-ray experiment. (c) Resonance spectra from the
two peaks recorded with $\pi$-polarized light in the region of
the La-$M_4$ and Ni-$L_3$ and $L_2$ resonances.}
\label{Kspace}
\end{figure}

At the Ni $(2p \rightarrow 3d)$ ($L_{2,3}$) resonance in the soft
x-ray range, we probed the $(2 \epsilon, 0 , 1)$ charge order and
the $(1-\epsilon, 0, 0)$ spin order peak, which, for the chosen
doping level, are both in the momentum space reachable at this
photon energy (see Fig.~\ref{Kspace}). At the Ni $(1s \rightarrow
4p)$ ($K$) resonance we probed the charge order at $\vec{q}=
(2-2\epsilon, 0 ,3)$.

\begin{figure}[t]
\includegraphics[clip,bb=58 86 463 738,width=6.5cm]{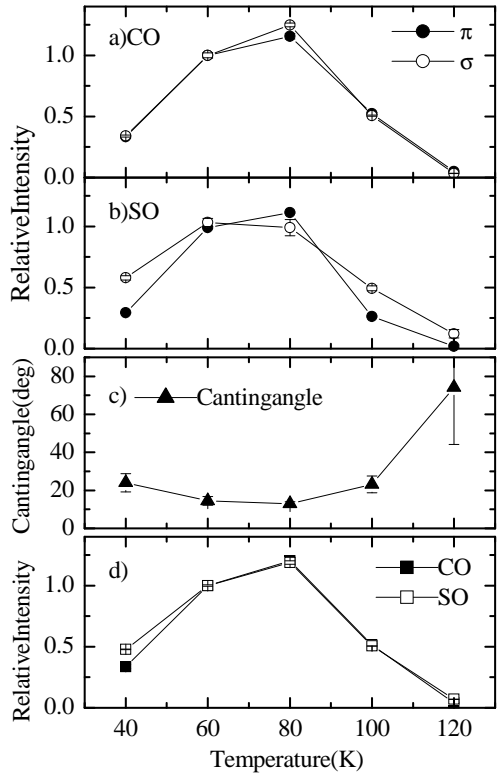}
\caption{Height of the maximum of the $L_3$ resonance above background
vs. temperature for a) the charge-order signal and b) the spin-order
signal. c) Canting angle as determined from the relative intensity of
the spin-order signals. d) Average of the CO signals from a) compared
to the total magnetic moment determined from the two SO curves.
For comparison the curves in a), b), and c) have been normalized at 60 K.
The lines are guides to the eye.}
\label{tpoldep}
\end{figure}

As a measure of the intensity of the charge and spin order peaks
we took the height of the maximum of the $L_3$ resonance
\cite{schuessler:05a} in a photon energy dependent scan with the
momentum transfer fixed at the corresponding peak position [see
Fig.~\ref{Kspace}(c)]. The obtained temperature dependence of the
CO signal is presented in Fig.~\ref{tpoldep}(a). The $\sigma$
signal is about 50 percent higher than the $\pi$ signal; for
comparison the data were normalized to the value at 60 K. The
effect of polarization on the temperature dependence is weak,
both curves almost match. The corresponding data for the SO
signal in Fig.~\ref{tpoldep}(b) were normalized in the same way.
The deviation between the SO temperature dependence observed with
the two light polarizations can be attributed to the variation of
the canting angle between the Ni spins and the $b$ direction. In
our scattering geometry with the detector at $\omega = 152^\circ$
we are probing with $\sigma$-polarized light essentially the
projection of the spins on the $a$ direction, while with $\pi$
polarized light we are sensitive to the $a$ and $b$ components of
the spin, the $b$ component weighted by a factor $\sin^2 \omega =
0.22$. If we assume the spins to be confined to the $ab$ plane
\cite{aeppli:88a,lee:01a} we find a temperature-dependent canting
angle as plotted in Fig.~\ref{tpoldep}(c). The overall behavior
of the spin direction is in agreement with results found for
other doping levels.\cite{lee:01a,freeman:02a,freeman:04a}

From the intensity of the magnetic signal in the two polarization
channels we can determine the \emph{total} spin signal. This is
plotted as the open symbols in Fig.~\ref{tpoldep}(d) in
comparison with the (averaged) CO signal. Both signals clearly
have a very similar temperature dependence in particular above 60
K. This finding is very different from the neutron results
presented in Fig.~\ref{neutron}.

\begin{figure}[!t]
\includegraphics[scale=1,clip,width=7.5cm]{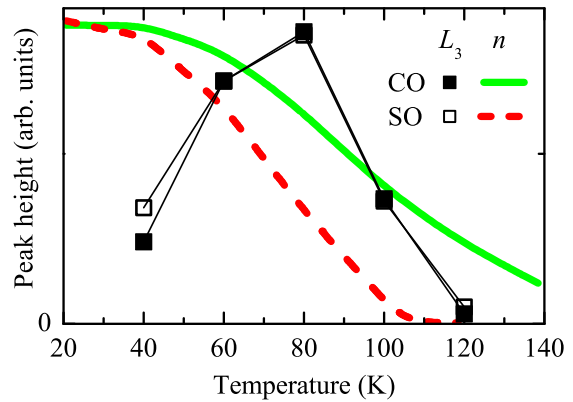}
\caption{(Color online) Comparison between neutron diffraction
(thick lines) and the RSXD results (symbols and thin lines) for
the temperature dependence of CO and SO.}
\label{neutron_and_l3}
\end{figure}

In Fig.~\ref{neutron_and_l3} we compare in more detail the
temperature dependences found in the neutron experiment (thick
lines) with the ones from Fig~\ref{tpoldep}(d). Focusing first on
the higher temperatures, both RSXD signals appear to decay on a
temperature scale that is in between those found by neutron
diffraction for CO and SO. Particularly dramatic are the
deviations between neutron and RSXD results at low temperatures:
While both neutron signals increase upon cooling and eventually
saturate, the RSXD signals decay upon cooling below 80 K. This
kind of different temperature dependence is in agreement with
what has been observed before using neutron
\cite{hayden:92a,sachan:95a,tranquada:95a,tranquada:96a,lee:97a,lee:01a}
and conventional x-ray diffraction,
\cite{du:00a,hatton:02a,hatton:02b,ghazi:04a,spencer:05a}
respectively.

\begin{figure}[!t]
\includegraphics[clip,bb=65 474 741 1053,width=8.5cm]{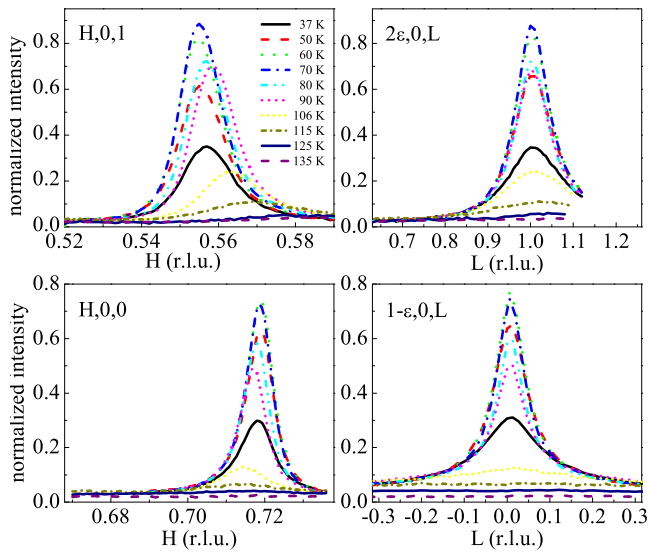}
\caption{(Color online) Scans through the superstructure peaks for different
temperatures along H (left) and along L (right). Upper: charge
order, lower: spin order.}
\label{width}
\end{figure}

\begin{figure}[!t]
\includegraphics[width=7.5cm]{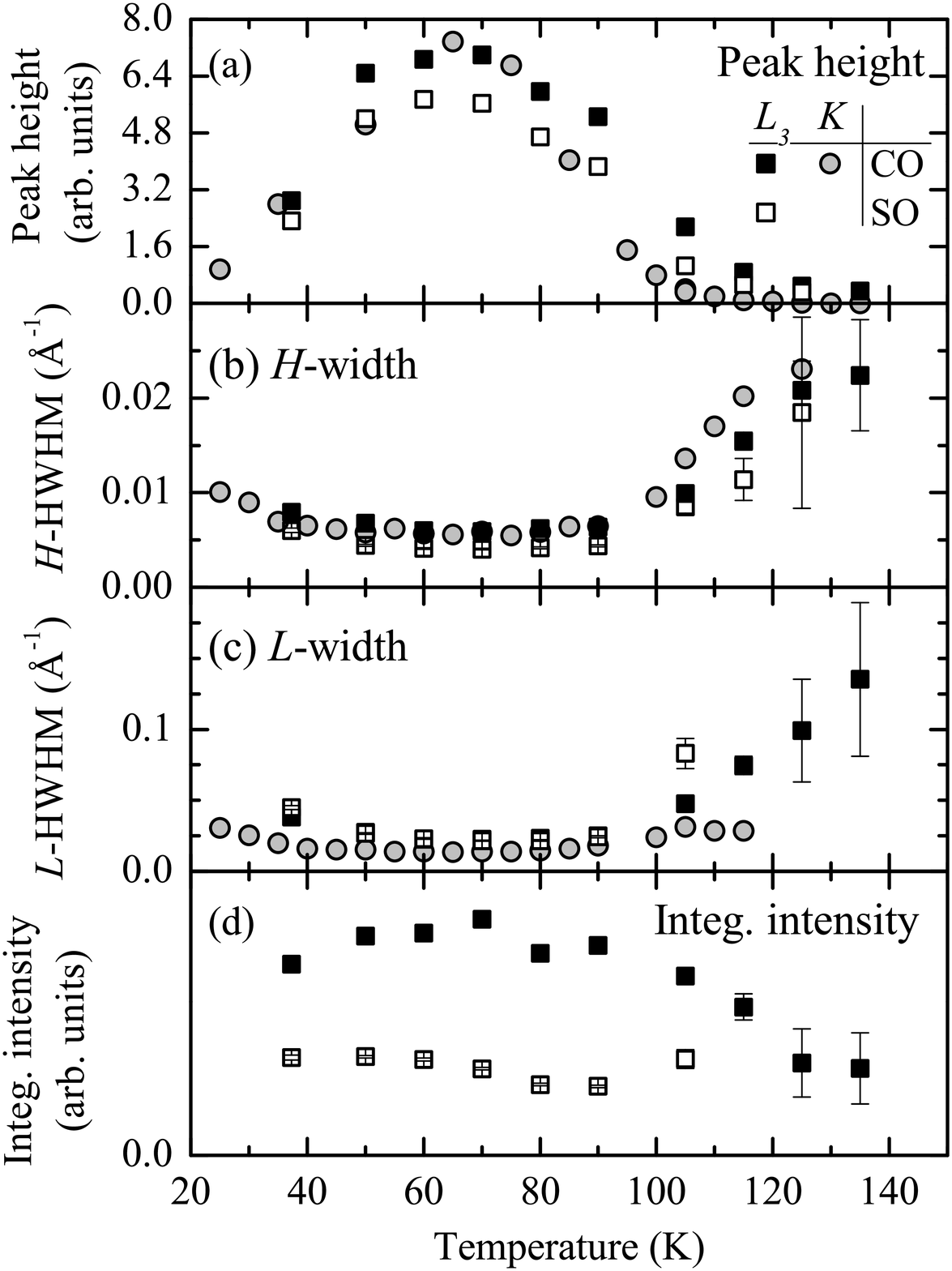}
\caption{Temperature dependence of the CO and SO peaks recorded
at the Ni-$L_3$ resonance (filled and open squares) and of the CO
peak recorded at the Ni-$K$ resonance (gray circles): (a) peak
height, (b) peak widths along $H$, (c) peak width along $L$, (d)
integrated intensity estimated from the $L_3$ data.}
\label{Int}
\end{figure}

In order to better characterize the temperature dependence we
recorded the RSXD data in a second more dense data set. We used
$\pi$-polarized light to probe the peak profile at the resonance
maximum using scans along $H$ and $L$ (Fig.~\ref{width}). The
temperature dependence of the peak height as shown in Fig.
\ref{Int}(a) is very similar to the one in Fig. \ref{tpoldep}
with the maximum occurring at 70 K in this case. Again CO and SO
signals behave similarly. We also show for comparison the CO
signal as obtained at the Ni-$K$ resonance (gray circles), which
shows a temperature dependence very similar to the soft x-ray
data.

As can be seen from the scans in Fig.~\ref{width} and from the
extracted peak widths (assuming a Lorentzian line shape) in
Fig.~\ref{Int} (b) and (c), a decreasing peak height upon heating
and cooling is accompanied by a broadening of the peak in both
directions of reciprocal space. Consequently the integrated peak
intensity is decaying much slower than the peak height. A similar
trend of peak broadening is also found at the $K$-edge. We tried
to estimate the integrated intensity from the $L_3$ data as the
product of the peak height and peak widths in $H$ and $L$
directions. We can do this without double counting of
intensities, because the $q$-space resolution broadening is about
10 times smaller than the smallest peak width in these two
directions. Along the third, $K$, direction the intensity is
mostly integrated because of the wide detector acceptance
perpendicular to the scattering plane. The result of our estimate
is shown in Fig. \ref{Int}(d). In particular upon cooling the
decrease of the peak heights seems to be fully compensated by the
increasing peak widths. Upon heating the integrated intensity of
the SO and CO signal stays almost constant up to 105 K. For
higher temperatures the spin-order signal is difficult to track,
because the temperature-dependent spin canting leads to an
additional intensity loss for $\pi$-polarization
[c.f.~Fig.~\ref{tpoldep}(b)]. The CO signal, however, is well
observable and we find a gradual decay only at higher
temperatures. We estimate the residual intensity at 135 K to 40
percent of the maximum intensity.

In a situation, where the peak height changes, but the integrated
intensity is conserved, the detected signal depends strongly on
the $q$-space resolution of the experiment. An experiment
probing, e.g., CO, which is integrating over a wide range of $q$
space, will rather find a temperature dependence as shown in
Fig.~\ref{Int}(d): the signal is constant at low temperatures and
decays slowly only at high temperatures. On the other hand, an
experiment with high $q$-space resolution will mainly find a
decrease of the peak height with decreasing temperature. Since neutron diffraction experiments
are typically performed with lower resolution than x-ray
diffraction experiments, this effect can at least partially
explain the discrepancies between the low-temperature behavior
found by neutron diffraction and x-ray diffraction from this
system. The peak broadening can also explain, why the neutron CO
signal in Fig.~\ref{neutron_and_l3} appears to remain up to
higher temperatures than the x-ray signal: the peak height probed
in the x-ray experiments is decaying faster upon heating than the
(partially) integrated intensity probed by neutron diffraction.

Having established the importance of the different $q$-space
resolutions of the experiments, we are still left with the
following observation: the temperature dependence of the charge
and spin signals within the RSXD experiment is identical while it
is very dissimilar in neutron diffraction. This
observation cannot be attributed to differences in the $q$-space
resolutions since the peak broadening effect applies to both the
SO and CO signals with comparable widths as well. We need to look
for another explanation. We infer that one has to look to the
contribution from the \textit{inelastic} magnetic scattering to
the SO signal. For LSNO it is known, that the \emph{total}
magnetic scattering signal, i.e., the sum of elastic and
inelastic magnetic neutron scattering has about the same
temperature dependence as the CO signal.\cite{sachan:95a} The
RSXD experiment was performed without energy analysis of the
scattered photons. Unlike a neutron experiment, RSXD also lacks
an intrinsic energy analysis.\cite{footnote1} In that sense RXSD
is comparable to a hypothetical energy-integrated neutron
diffraction experiment,\cite{footnote2} and we have been probing
the sum of static and fluctuating magnetic correlations.

To conclude, we studied the temperature dependence of SO and CO
using RSXD, neutron diffraction and Ni-$K$-edge RXD for the CO
signal. Differences in the $q$-space resolution of the different
techniques have to be taken into account for a proper evaluation
of diffraction intensities associated with the SO and CO
superstructures. We find that in RSXD the SO and CO decay on the
same temperature scale upon heating, very different from that in
(elastic) neutron diffraction. The reason for this difference is
that RSXD is probing the static and the fluctuating magnetic
order and is hence in fact comparable to an energy-integrated
neutron diffraction experiment. This in fact may open up an
interesting experimental possibility to search for fluctuating order
in other systems by comparing resonant soft x-ray diffraction and
neutron diffraction data.

\begin{acknowledgments}

We gratefully acknowledge the expert support and excellent
working conditions at BESSY and thank E. Weschke for making his
UHV diffractometer available for this work. We thank L. Hamdan and
the mechanical workshop of the II. Physikalische Institut in
Cologne for their skillful technical assistance, and T. Koethe
for help in preparing the experiment. Preliminary experiments
were carried out at X1B of the NSLS and we acknowledge
experimental support by P. Abbamonte and A. Rusydi. The research
in K\"oln is supported by the Deutsche Forschungsgemeinschaft
through SFB 608. Work at BESSY was supported by the BMBF through
project 05 ES3XBA/5.

\end{acknowledgments}


\begin{thebibliography}{28}
\bibitem[{foo({\natexlab{a}})}]{footnote1}
\bibinfo{howpublished}{In neutron scattering the energy of magnetic
  fluctuations of the order of 10 meV leads to a sizable change of the neutron
  momentum $k_\mathrm{Neutron} \propto {E_\mathrm{Neutron}}^{1/2}$ due to the
  small neutron energies, which are of the order of 50 meV. For soft x-rays
  with photon energies, $E_\mathrm{Photon}$, of several 100 eV, such an energy
  change has only negligible influence on the photon wave vector,
  $k_\mathrm{Photon} \propto E_\mathrm{Photon}$.}
\bibitem[{foo({\natexlab{b}})}]{footnote2}
\bibinfo{howpublished}{In general such an energy integrating neutron scattering
  experiment is not possible directly, because even without energy analyzer a
  neutron diffraction experiment does not integrate all the inelastic
  scattering for a given scattering vector $\vec{q}$. For layered materials
  like LSNO with two-dimensional magnetic fluctuations, however, one can set
  the direction of the outgoing neutron momentum parallel to the
  $c^*$-direction, i.e. perpendicular to the plane of magnetic correlations and
  in that way achieve an energy integration.}
\end{thebibliography}
\end{document}